\documentclass[aps,prl,preprint,groupedaddress]{revtex4}
\usepackage{graphicx}
\usepackage{dcolumn}
\usepackage{bm}
\usepackage{hyperref}
\hypersetup{colorlinks,bookmarksopen,bookmarksnumbered,citecolor=blue,pdfstartview=FitH}
\begin{document}
\preprint{YITP-SB-05-22}

\title{$N=4$ Super Yang-Mills NMHV Loop Amplitude in Superspace}

\author{Yu-tin Huang }
 \email{yhuang@grad.physics.sunysb.edu}\affiliation{State University of New York YITP, Stony Brook}

\date{\today}

\begin{abstract}
Here we construct N=4 SuperYang-Mills 6 point NMHV loop amplitude
(amplitudes with three minus helicities) as a full superspace
form, using the $SU(4)_{R}$ anti-commuting spinor variables.
Amplitudes with different external particle and cyclic helicity
ordering are then just a particular expansion of this fermionic
variable. We've verified this by explicit expansion obtaining
amplitudes with two gluinos calculated before. We give results for
all gluino
$A(\Lambda^{-}\Lambda^{-}\Lambda^{-}\Lambda^{+}\Lambda^{+}\Lambda^{+})$and
all scalar $A(\phi\phi\phi\phi\phi\phi)$scattering amplitudes. A
discussion of using MHV vertex approach to obtain these amplitudes
are given, which implies a possible simplification for general
loop amplitudes.
\end{abstract}
\maketitle
\section{1.Introduction}

N=4 SuperYang-Mills one loop amplitudes have the special property
of being cut constructible \cite{1}\cite{2}, that is they are
uniquely determined by their unitary cuts. It was shown in
\cite{1} that loop amplitudes can be written as a combination of
scalar box integrals
 (for definition of the scalar box integral and scalar box function
see appendix for \cite{1}) with rational coefficients. Therefore
the calculation of one loop amplitude is reduced to determining
the coefficients in front of these box integrals, which is done by
analyzing the cuts of the amplitude and matching them with the
cuts of the scalar box integrals. Unfortunately complication
arises from the fact that some of the cuts are shared by more than
one box integral. This was dramatically simplified in \cite{3} by
using generalized unitarity (quadruple cuts) cuts \cite{4} to
analyze the leading singularities which turns out to be unique in
the box integrals.

Recently \cite{5} it was shown by using Supersymmetric Ward
Identity (SWI)\cite{6} one can derive N $=4$ SYM NMHV 6 point tree
and loop amplitudes with gluinos or scalars from
 their pure gluonic partners. Since SWI corresponds to a
 transformation in superspace, one would guess this implies the
 existence of a full superspace amplitude while amplitudes with
 different external particle species are considered as different
 component of the superspace expansion.
The simplest superspace amplitude was the MHV and $\overline{MHV}$
tree amplitude written down by Nair \cite{7}. Since the work by
Witten \cite{8} which showed that perturbative N=4 SYM is dual to
a particular string theory with super-twistor space as its target
space, various new techniques have been developed to calculate N=4
SYM amplitudes more efficiently\cite{9}\cite{10}. The MHV vertex
construction \cite{10}, which uses MHV vertices as the basic
building block of the scattering amplitudes, provides a convenient
method to construct the amplitudes in superspace form. This was
done for the NMHV tree amplitude in \cite{11}. At loop level the
valedictory of MHV vertex approach was proven to give the same
result as that in field theory in \cite{12} for MHV
 loop amplitudes, and \cite{13} reproduces the relationship between the color leading amplitudes and sub-leading amplitudes. At this point it is
 natural to continue with MHV vertices to compute NMHV loop which
 would require three MHV vertices connected by three propagators, and this should give the full superspace form of the NMHV loop.
At this point it is not clear how the correct scalar box functions
[1] should arise in this formalism. One of the complication is for
more than two fermionic delta functions (there is one for each MHV
vertex), after the expansion in superspace there will be multiple
spinor products that contain the off shell continuation spinor of
the propagator, which takes different form with different external
particle specie. Since these spinor products should be integrated
over, the integrand for the gluonic amplitudes will be
dramatically different from the ones with gluinos, implying one
can only derive the box functions from the superspace expansion
one term at a time and not in the original superspace full form.

In \cite{5} the SWI identities were not used directly upon the
 coefficients in front of the box integrals for the gluonic amplitude, but rather the
 coefficients in front of a particular combination of box integrals, which originated from the three different cut channels \cite{2}. To realize
 the superspace amplitude all one needs to observe is that for
 the six point amplitude the three channels from which the cuts
 were computed, the tree graphs on either side of the cuts
 always come in MHV and $\overline{MHV}$ pair. Since MHV and
 $\overline{MHV}$
 tree can be written straight forwardly in superspace form, one
 naturally derives the six point one loop NMHV amplitude for all
 helicity configuration and external species as one superspace amplitude
by fusing the two tree amplitude. In the following we present the
amplitude in its full superspace form and confirm our result by
explicitly expanding out the terms that give the correct
amplitudes with two gluino obtained in \cite{5}. We will also give
a brief demonstration of how one could obtain the field theory
result for the loop amplitude from the MHV vertex approach (CSW)
\cite{10}.

\section{2. The Construction}

The n point MHV and $\overline{MHV}$ tree level amplitudes have a
remarkable simple form. For MHV tree \cite{7}:

\begin{equation}\label{eq.1}
A(...j^{-}.....i^{-}...)_{tree}=\frac{\delta^{8}(\sum_{i=1}^{n}\lambda_{i}\eta_{i}^{A})}{\Pi^{n}_{i=1}<ii+1>}
\end{equation}
where
\begin{equation}\label{eq.1}
\delta^{8}(\sum_{i=1}^{n}\lambda_{i}\eta_{i}^{A})=\frac{1}{2}\prod_{A=1}^{4}(\sum_{i=1}^{n}\lambda_{i}^{\alpha}\eta_{i}^{A})(\sum_{i=1}^{6}\lambda_{i\alpha}\eta_{i}^{A})
\end{equation}
as for $\overline{MHV}$ tree:
  \begin{equation}\label{eq.1}
A(...j^{+}.....i^{+}...)=\frac{\delta^{8}(\sum_{i=1}^{6}\widetilde{\lambda_{i}}\widetilde{\eta_{i}^{A})}}{\Pi^{6}_{i=1}[ii+1]}
\end{equation}
Here we've omitted the energy momentum conserving delta function
and the group theory factor. After expansion in the fermionic
parameters $\eta_{i}^{A}$, one can obtain MHV amplitudes with
different helicity ordering ($++---,+-+--$...etc) and different
particle content.

We proceed to construct the full N=4 SYM NMHV 1-loop six point
amplitudes by following the original gluonic calculation \cite{2},
where the amplitude was computed from the cuts of the three
channels $t_{123}$ $t_{234}$ $t_{345}$
$(t_{ijk}=(k_{i}+k_{j}+k_{l})^{2})$, except now the tree
amplitudes across the cuts are written in supersymmetric form. We
find that the propagator momentum integrals from which the various
scalar box functions arise are the same for different external
particles. Thus with the gluon amplitude already computed all we
need to do is extract away the part of the gluon coefficient that
came from the expansion of the two fermionic delta function, the
remaining pre factor will be universal and has its origin from the
denominator of eq.(1) and (3). The N=4 SYM 6 point NMHV loop
amplitude for the gluonic case was given \cite{2} as
\begin{equation}\label{eq.1}
A(...j^{-}.....i^{-}...)_{loop}=c_{\Gamma}[B_{1}W_{6}^{(1)}+B_{2}W_{6}^{(2)}+B_{3}W_{6}^{(3)}]
\end{equation}
where $W_{6}^{(i)}$ contains particular combination of the
two-mass-hard and one-mass box functions \cite{1}. The full 6
point
 NMHV loop amplitude for any given set of external particle and
helicity ordering are then given with the following coefficients :
\begin{eqnarray}\label{eq.1}
B_{1}=\frac{\delta^{8}(\sum_{i=1}^{3}\widetilde{\lambda_{i}}\widetilde{\eta_{i}}-\widetilde{l_{1}}\widetilde{\eta_{1}}+\widetilde{l_{2}}\widetilde{\eta_{2}})\delta^{8}(\sum_{i=4}^{6}\lambda_{i}\eta_{i}-l_{2}\eta_{2}+l_{1}\eta_{1})}{t_{123}}B_{0}\\
\nonumber+\frac{\delta^{8}(\sum_{i=1}^{3}\lambda_{i}\eta_{i}-l_{1}\eta_{1}+l_{2}\eta_{2})\delta^{8}(\sum_{i=4}^{6}\widetilde{\lambda_{i}}\widetilde{\eta_{i}}-\widetilde{l_{2}}\widetilde{\eta_{2}}+\widetilde{l_{1}}\widetilde{\eta_{1}})}{t_{123}}B_{0}^{\dag}
\end{eqnarray}
\begin{eqnarray}\label{eq.1}
B_{2}=\frac{\delta^{8}(\sum_{i=2}^{4}\widetilde{\lambda_{i}}\widetilde{\eta_{i}}-\widetilde{l_{1}}\widetilde{\eta_{1}}+\widetilde{l_{2}}\widetilde{\eta_{2}})\delta^{8}(\sum_{i=5}^{1}\lambda_{i}\eta_{i}-l_{2}\eta_{2}+l_{1}\eta_{1})}{t_{234}}B_{+}
\\
\nonumber+\frac{\delta^{8}(\sum_{i=2}^{4}\lambda_{i}\eta_{i}-l_{1}\eta_{1}+l_{2}\eta_{2})\delta^{8}(\sum_{i=5}^{1}\widetilde{\lambda_{i}}\widetilde{\eta_{i}}-\widetilde{l_{2}}\widetilde{\eta_{2}}+\widetilde{l_{1}}\widetilde{\eta_{1}})}{t_{234}}B_{+}^{\dagger}
\end{eqnarray}
\begin{eqnarray}\label{eq.1}
B_{3}=\frac{\delta^{8}(\sum_{i=3}^{5}\widetilde{\lambda_{i}}\widetilde{\eta_{i}}-\widetilde{l_{1}}\widetilde{\eta_{1}}+\widetilde{l_{2}}\widetilde{\eta_{2}})\delta^{8}(\sum_{i=6}^{2}\lambda_{i}\eta_{i}-l_{2}\eta_{2}+l_{1}\eta_{1})}{t_{345}}B_{-}\\
\nonumber+\frac{\delta^{8}(\sum_{i=3}^{5}\lambda_{i}\eta_{i}-l_{1}\eta_{1}+l_{2}\eta_{2})\delta^{8}(\sum_{i=6}^{2}\widetilde{\lambda_{i}}\widetilde{\eta_{i}}-\widetilde{l_{2}}\widetilde{\eta_{2}}+\widetilde{l_{1}}\widetilde{\eta_{1}})}{t_{345}}B_{-}^{\dagger}
\end{eqnarray}
where we define :
\begin{equation}
B_{0}=i\frac{1}{[12][23]<45><56><1|K_{123}|4><3|K_{123}|6>}
\end{equation}
and
\begin{equation}
B_{+}=B_{0}|_{j\rightarrow j+1}\;B_{-}=B_{0}|_{j\rightarrow j-1}
\end{equation}
with $<A|K_{ijk}|B>=[Ai]\langle iB\rangle+[Aj]\langle
jB\rangle+[Ak]\langle kB\rangle$. Each coefficient is expressed in
two terms, this corresponds to the assignment of helicity for the
propagators $l_{1}$ and $l_{2}$ which for specific assignments
will reverse the MHV and $\overline{MHV}$ nature of the two tree
amplitude across the cut (fig-1). The presence of the loop momenta
seems perplexing at this point since all loop momenta should have
been integrated out to give the box functions. As we will see on a
case by case basis this comes as a blessing. The actual expansion
for a particular set of helicity ordering and external particles
contains multiple terms, the presence of loop momentum forces one
to regroup the terms such that the loop momentum forms kinematic
invariants, it is after this regrouping that one obtains previous
known results.

The amplitudes for different external particles are computed as an
expansion in the $SU(4)_{R}$ anti-commuting fermionic variables
$\eta$. Choosing particular combinations following \cite{14}
\begin{figure}
\includegraphics{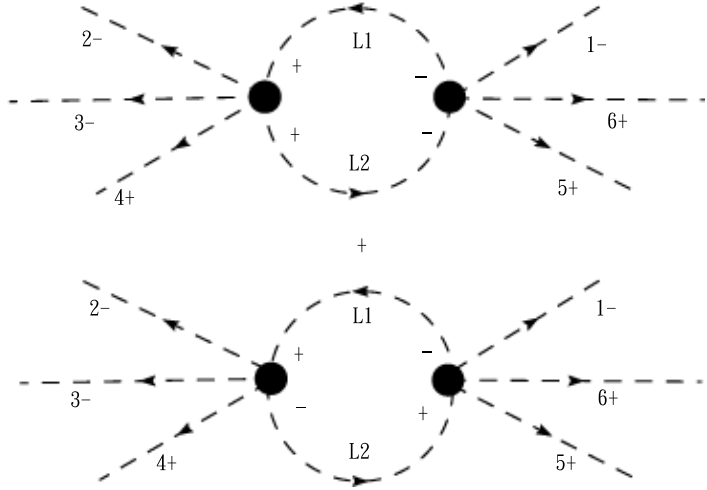}
\caption{\label{fig:2} Here we show for a particular case of the
gluonic NMHV loop amplitude, different assignment of helicity for
the propagators will change the MHV or $\overline{MHV}$ nature of
each vertex. In the upper graph the propagators has the same
helicity on each vertex while in the lower they are opposite. This
result in a MHV vertex on the left for the upper graph and an
$\overline{MHV}$ vertex on the left for the lower graph. In
practice one has to sum these two possibilities which is the
reason we have two terms in eq.(5)-(7)}
\end{figure}
\begin{eqnarray}\label{eq.1}
g_{i}^{-}=\eta^{1}_{i}\eta^{2}_{i}\eta^{3}_{i}\eta^{4}_{i},
\;\phi^{AB}_{i}=\eta^{A}_{i}\eta^{B}_{i},
\;\Lambda^{1-}_{i}=-\eta^{2}_{i}\eta^{3}_{i}\eta^{4}_{i},\;\Lambda^{2-}_{i}=-\eta^{1}_{i}\eta^{3}_{i}\eta^{4}_{i}\\
\nonumber
\qquad\Lambda^{3-}_{i}=-\eta^{1}_{i}\eta^{2}_{i}\eta^{4}_{i},
\qquad\Lambda^{4-}_{i}=-\eta^{1}_{i}\eta^{2}_{i}\eta^{3}_{i},\qquad\Lambda^{A+}_{i}=\eta^{A}_{i},\qquad
g_{i}^{+}=1
\end{eqnarray}
The superscript represents which flavor the particle carries, in
the N=4 multiplet there are four gluinos and six scalars.
Corresponding combination in the $\widetilde{\eta}$ follows:
\begin{eqnarray}\label{eq.1}
g_{i}^{+}=\widetilde{\eta^{1}_{i}}\widetilde{\eta^{2}_{i}}\widetilde{\eta^{3}_{i}}\widetilde{\eta^{4}_{i}},
\;\phi^{AB}_{i}=\widetilde{\eta^{C}_{i}}\widetilde{\eta^{C}_{i}},
\;\Lambda^{1+}_{i}=-\widetilde{\eta^{2}_{i}}\widetilde{\eta^{3}_{i}}\widetilde{\eta^{4}_{i}},\;\Lambda^{2+}_{i}=-\widetilde{\eta^{1}_{i}}\widetilde{\eta^{3}_{i}}\widetilde{\eta^{4}_{i}}\\
\nonumber
\qquad\Lambda^{3+}_{i}=-\widetilde{\eta^{1}_{i}}\widetilde{\eta^{2}_{i}}\widetilde{\eta^{4}_{i}},
\qquad\Lambda^{4+}_{i}=-\widetilde{\eta^{1}_{i}}\widetilde{\eta^{2}_{i}}\widetilde{\eta^{3}_{i}},\qquad\Lambda^{A-}_{i}=\widetilde{\eta^{A}_{i}},\qquad
g^{-}_{i}=1
\end{eqnarray}
Thus a particular term in the expansion corresponds to a
particular assignment of the fermionic variables to the external
particle and results in an amplitude with a particular set of
external particle specie and helicity ordering. In the next two
section we show by expanding eq.(5),(6),(7) and following the
above dictionary one can recover the amplitudes containing two
 same color gluino with different helicity ordering computed in \cite{5}.

\subsection{2.1 $B_{1}$ Coefficient $\Rightarrow$ $t_{123}$ cut}
First we look at the $t_{123}$ cut which correspond to the $B_{1}$
coefficient. For the purely gluonic amplitude
$A(g_{1}^{-}g_{2}^{-}g_{3}^{-}|g_{4}^{+}g_{5}^{+}g_{6}^{+})$
 (we use a bar to indicate the cut ), we have
only one particle assignment for the loop propagators:
\begin{equation}
l_{1}=g^{+}\;,l_{2}=g^{+}
\end{equation}
Here the assignment of helicity is labelled with respect to the
$\overline{MHV}$ vertex. Therefore we get only contribution from
the first term in eq.(5), the expansion from the delta function
gives
$\langle\l_{1}l_{2}\rangle^{4}[l_{1}l_{2}]^{4}=(l_{1}-l_{2})^{8}=t_{123}^{4}$
and therefore $B_{1}=t_{123}^{3}B_{0}$ which matches eq.(5.4) in
\cite{5}.

For the two gluino amplitudes first we look at
$A(\Lambda^{-}_{1}g_{2}^{-}g_{3}^{-}|\Lambda^{+}_{4}g_{5}^{+}g_{6}^{+})$
from the delta function expansion the we have helicity assignments
:
\begin{equation}
l_{1}=\Lambda^{+}\;l_{2}=g^{+}\;,\;+(\;exchage\;between\;
l_{1}\;and\; l_{2}\;)
\end{equation}
Again only the first term in eq.(5) gives contribution :
\begin{equation}\label{eq.1}\langle l_{1}l_{2}\rangle^{3}[l_{1}l_{2}]^{3}([1l_{1}]\langle l_{1}4\rangle-[1l_{2}]\langle
l_{2}4\rangle)=t_{123}^{3}\langle 1|K_{123}|4\rangle
\end{equation}
Note that only when the external gluino carry the same flavor will
this term contribute. Since in \cite{5} the two gluino amplitude
was derived using N=1 SWI, the two gluinos carry the same flavor.
Thus we have
\begin{equation}
B_{1}(\Lambda^{-}_{1}g_{2}^{-}g_{3}^{-}|\Lambda^{+}_{4}g_{5}^{+}g_{6}^{+})=i\frac{t_{123}^{2}\langle
1|K_{123}|4\rangle}{[12][23]<45><56><1|K_{123}|4><3|K_{123}|6>}
\end{equation}
This is exactly the result of \cite{5}. Other non-cyclic
permutations of two gluino amplitude calculated in \cite{5} at
this cut do not change the assignment of the propagators thus the
amplitude remains the same form apart from the labelling of the
position of the two gluinos.

\subsection{\label{sec:level2}2.2 $B_{2}$ Coefficient $\Rightarrow$$t_{234}$ cut}
For this cut with different helicity assignment of the
propagators, contribution can arise from both terms. Propagators
with the same helicities (here we mean they are both plus or minus
regardless of the specie) get its contribution from one term while
the rest from the other, this is why $B_{2}$ was split in two
terms in the original computation of the gluon amplitude \cite{2}.
We deal with the same helicity first since there is only one way
of assigning propagators.

$B_{2}(\Lambda^{-}_{1}|g_{2}^{-}g_{3}^{-}\Lambda^{+}_{4}|g_{5}^{+}g_{6}^{+})_{same
helicity}=0$ since there is no way of assigning same helicity
particles to the propagators.

For
$B_{2}(g_{1}^{-}|\Lambda^{-}_{2}g_{3}^{-}\Lambda^{+}_{4}|g_{5}^{+}g_{6}^{+})_{same
helicity}$ we have
\begin{equation}\label{eq.1} l_{1}=g^{-}\;
l_{2}=g^{-}
\end{equation}
this receives contribution from the second term in eq.(6) which is
$\langle23\rangle^{3}\langle43\rangle[56]^{4}$ thus giving
\begin{equation}\label{eq.1} B_{2}(g_{1}^{-}|\Lambda^{-}_{2}g_{3}^{-}\Lambda^{+}_{4}|g_{5}^{+}g_{6}^{+})_{same
helicity}=(\frac{\langle23\rangle^{3}\langle43\rangle[56]^{4}}{t_{234}})B_{+}^{\dag}
\end{equation}
For
$B_{2}(g_{1}^{-}|\Lambda^{-}_{2}g_{3}^{-}g_{4}^{+}|\Lambda^{+}_{5}g_{6}^{+})_{same
helicity}$ we have
\begin{equation}\label{eq.1} l_{1}=g^{-}\;
l_{2}=\Lambda^{-}\;,+\;(exchange\;between \;l_{1},l_{2})
\end{equation}
This gives contribution
$\langle23\rangle^{3}[56]^{3}(\langle3l_{1}\rangle[l_{1}6]-\langle3l_{2}\rangle[l_{2}6])=\langle23\rangle^{3}[56]^{3}\langle3|K_{234}|6\rangle$
giving
\begin{equation}\label{eq.1}
B_{2}(g_{1}^{-}|\Lambda^{-}_{2}g_{3}^{-}g_{4}^{+}|\Lambda^{+}_{5}g_{6}^{+})_{same
helicity}=(\frac{\langle23\rangle^{3}[56]^{3}\langle3|K_{234}|6\rangle}{t_{234}})B_{+}^{\dag}
\end{equation}
Now we move to configurations with different helicity. For
$B_{2}(\Lambda^{-}_{1}|g_{2}^{-}g_{3}^{-}\Lambda^{+}_{4}|g_{5}^{+}g_{6}^{+})_{Diff
Helicity}$ we have :
\begin{equation}\label{eq.1} l_{1}=\Lambda^{+}\;
l_{2}=g^{-} \;, l_{1}=\Lambda^{-}\; l_{2}=\phi
\;,+\;(exchange\;between \;l_{1},l_{2})
\end{equation}
For fix flavored $\Lambda_{4}^{+}$ and $\Lambda_{1}^{-}$ we have
to sum up all possible flavors for the internal gluino. This gives
a contribution of
 \begin{eqnarray}\label{eq.1}
[1l_{1}]^{3}[l_{1}l_{2}]\langle 4l_{2}\rangle^{3}\langle
l_{1}l_{2}\rangle-3[l_{1}l_{2}][l_{2}4][l_{1}4]^{2}\langle
l_{1}l_{2}\rangle\langle l_{2}1\rangle\langle
l_{1}1\rangle^{2}\\
\nonumber+3[l_{1}l_{2}][l_{1}4][l_{2}4]^{2}
 \langle l_{1}l_{2}\rangle\langle l_{1}1\rangle\langle
l_{2}1\rangle^{2}\\
\nonumber-[1l_{2}]^{3}[l_{1}l_{2}]\langle 4l_{2}\rangle^{3}\langle
l_{1}l_{2}\rangle=t_{123}(\langle1|l_{1}-l_{2}|4\rangle)^{3}=t_{123}\langle1|K_{123}|4\rangle^{3}
\end{eqnarray}
Therefore
\begin{equation}\label{eq.1}
B_{2}(\Lambda^{-}_{1}|g_{2}^{-}g_{3}^{-}\Lambda^{+}_{4}|g_{5}^{+}g_{6}^{+})_{DiffHelicity}=\frac{\langle1|K_{123}|4\rangle^{3}}{t_{123}^{3}}B_{+}
\end{equation}
For
$B_{2}(g_{1}^{-}|\Lambda^{-}_{2}g_{3}^{-}\Lambda^{+}_{4}|g_{5}^{+}g_{6}^{+})_{Diff
Helicity}$ we have:
\begin{equation}\label{eq.1}
l_{1}=g^{-}\; l_{2}=g^{+} \;, l_{1}=\Lambda^{+}\;
l_{2}=\Lambda^{-} \;,l_{1}=\phi\; l_{2}=\phi
\;,\;+\;(exchange\;between \;l_{1},l_{2})
\end{equation}
Here whether or not $\Lambda_{4}^{+}$ and $\Lambda_{1}^{-}$ carry
the same flavor will effect the number of ways one can assign
flavor to the internal gluino and scalar. For the same flavor we
have
\begin{equation}\label{eq.1} -([4l_{1}]\langle
l_{1}1\rangle-[4l_{2}]\langle l_{2}1\rangle)^{3}([2l_{1}]\langle
l_{1}1\rangle-[2l_{2}]\langle l_{2}1\rangle)=-(\langle
4|K_{234}|1\rangle)^{3}(\langle 2|K_{234}|1\rangle)
\end{equation}
Thus
\begin{equation}
B_{2}(g_{1}^{-}|\Lambda^{-}_{2}g_{3}^{-}\Lambda^{+}_{4}|g_{5}^{+}g_{6}^{+})_{Diff
Helicity}=(\frac{-(\langle 4|K_{234}|1\rangle)^{3}(\langle
2|K_{234}|1\rangle)}{t_{234}})B_{+} \end{equation} For
$B_{2}(g^{-}_{1}|\Lambda_{2}^{-}g_{3}^{-}g_{4}^{+}|\Lambda^{+}_{4}g^{+}_{6})_{Diffhelicity}$
we have :
\begin{equation}\label{eq.1}
l_{1}=g^{-}\; l_{2}=\Lambda^{+} \;, l_{1}=\Lambda^{-}\; l_{2}=\phi
\;,\;+\;(exchange\;between \;l_{1},l_{2})
\end{equation}
This gives contribution :
\begin{eqnarray}
-\langle1l_{1}\rangle^{3}\langle15\rangle[l_{1}4]^{3}[42]+3\langle1l_{1}\rangle^{2}[l_{1}4]^{2}\langle1l_{2}\rangle[l_{2}4][42]\langle15\rangle
\\
\nonumber-3\langle1l_{2}\rangle^{2}[l_{2}4]^{2}\langle1l_{1}\rangle[l_{1}4][42]\langle15\rangle+\langle1l_{1}\rangle^{3}\langle15\rangle[l_{1}4]^{3}[42]\\
\nonumber=-(\langle4|K_{234}|1\rangle)^{3}[42]\langle15\rangle
\end{eqnarray}
Thus
\begin{equation}
B_{2}(g^{-}_{1}|\Lambda_{2}^{-}g_{3}^{-}g_{4}^{+}|\Lambda^{+}_{4}g^{+}_{6})_{Diffhelicity}=\frac{-(\langle4|K_{234}|1\rangle)^{3}[42]\langle15\rangle}{t_{234}}B_{+}
\end{equation}
 Adding eq.(17),(19),(22),(25) and (28) together gives the $B_{2}$ coefficient of the gluino anti-gluino
pair amplitudes computed in \cite{5}. Coefficients for the next
cut can be calculated in similar way, we've checked it gives the
same result as that derived in \cite{5}.

  It is straight forward to compute amplitudes that
involve more than one pair of gluino or scalar. The new amplitudes
are :
\begin{eqnarray}\label{eq.1}
A(g^{-}g^{+}\Lambda^{+}\Lambda^{-}\Lambda^{-}\Lambda^{+})\;,A(g^{-}g^{+}\phi\phi\phi\phi)\;,A(\phi\phi\phi\phi\phi\phi)\;A(\Lambda^{-}\Lambda^{-}\Lambda^{-}\Lambda^{+}\Lambda^{+}\Lambda^{+})\;,\\
\nonumber
A(\Lambda^{-}\Lambda^{+}\phi\phi\phi\phi)\;,A(\Lambda^{-}\Lambda^{-}\Lambda^{+}\Lambda^{+}\phi\phi)\;,A(\Lambda^{-}\Lambda^{+}\phi
g^{-}g^{+}g^{+})
\end{eqnarray}
Complication arises for these amplitudes because non-gluon
particles carry less superspace variables and increase the amount
of spinor combination. Luckily with the specification of the
flavor for the external particles, the propagators are restricted
to take certain species. This is discussed in detail in the next
section where we calculate the all gluino and all scalar
amplitude.
\section{3. Amplitudes with all gluinos and all scalars }
Here we present N=4 SYM NMHV loop amplitudes with all gluino and
all scalars. These amplitudes were derived from explicit expansion
of eq.(5)-(7). Since scalars and gluinos carry less fermionic
parameters as seen in eq.(10)(11), the spinor product that arises
from the fermionic delta function becomes complicated. The final
coefficient should not contain the off shell propagator spinor,
thus one can use this as a guideline to group the spinor products
to form kinematic invariant terms. With specific flavors this also
restrict the possible species for propagators.
\subsection{\label{sec:level2}3.1
$A(\Lambda_{1}^{\emph{1}+}\Lambda_{2}^{\emph{2}+}\Lambda_{3}^{\emph{3}+}\Lambda_{4}^{\emph{1}-}\Lambda_{5}^{\emph{2}-}\Lambda_{6}^{\emph{3}-})$}For
the six gluino amplitude we look at amplitudes with all three
positive helicity gluino carrying different flavor. The negative
helicities also carry different flavor and is the same set as the
positive. For $t_{123}$ the flavors of the internal particles are
uniquely determined.
\begin{equation}
l_{1}=\Lambda^{-}\;l_{2}=g^{+}\;,\;l_{1}=\Lambda^{+}\;l_{2}=\phi\;,\;+exchange
\end{equation}
This gives
\begin{eqnarray}
B_{1}(\Lambda_{1}^{\emph{1}+}\Lambda_{2}^{\emph{2}+}\Lambda_{3}^{\emph{3}+}|\Lambda_{4}^{\emph{1}-}\Lambda_{5}^{\emph{2}-}\Lambda_{6}^{\emph{3}-})=(\langle1|K_{123}|5\rangle\langle2|K_{123}|6\rangle\langle3|K_{123}|4\rangle\\
\nonumber+\langle1|K_{123}|4\rangle\langle2|K_{123}|5\rangle\langle3|K_{123}|6\rangle+\langle1|K_{123}|6\rangle\langle2|K_{123}|4\rangle\langle3|K_{123}|5\rangle)B_{0}^{\dag}
\end{eqnarray}
Next we look at $t_{345}$ cut. The propagator assignment with same
helicity (the definition of same or different helicity again
follows that of the previous paragraph ) would be :
\begin{equation}
l_{1}=g^{-}\;l_{2}=\Lambda^{-}\;,\;+ exchange\;propagator
\end{equation}
this gives
\begin{eqnarray}
B_{3}(\Lambda_{1}^{\emph{1}+}\Lambda_{2}^{\emph{2}+}|\Lambda_{3}^{\emph{\emph{3}}+}\Lambda_{4}^{\emph{1}-}\Lambda_{5}^{\emph{2}-}|\Lambda_{6}^{\emph{3}-})_{SameHelicity}=\langle
45\rangle^{2}[12]^{2}\{\langle34\rangle[61]\langle5|K_{345}|2\rangle\\
\nonumber+\langle34\rangle[62]\langle5|K_{345}|1\rangle
+\langle35\rangle[61]\langle4|K_{345}|2\rangle+\langle35\rangle[62]\langle4|K_{345}|1\rangle\}B_{-}^{\dag}
\end{eqnarray}
There are two ways of assigning different helicity propagators
\begin{equation}
l_{1}=g^{-}\;l_{2}=\Lambda^{+}\;,\;or
l_{1}=\Lambda^{-}\;l_{2}=\phi\;,\;+\;exchange
\end{equation}
Note however for the present set of flavors, there is no
consistent way of assigning flavors when the propagators are a
gluon and a gluino. Thus we are left with the gluino scalar
 possibility with it's flavor uniquely determined.
\begin{eqnarray}
B_{3}(\Lambda_{1}^{\emph{1}+}\Lambda_{2}^{\emph{2}+}|\Lambda_{3}^{\emph{3}+}\Lambda_{4}^{\emph{1}-}\Lambda_{5}^{\emph{2}-}|\Lambda_{6}^{\emph{3}-})_{DiffHelicity}=\langle16\rangle\langle62\rangle[43][35]\langle6|K_{345}|3\rangle
t_{345}B_{-}
\end{eqnarray}

Luckily there is no need to compute $B_{2}$ coefficients since it
is related to $B_{3}$ by symmetry.
\subsection{\label{sec:level2}3.2 $A(\phi_{1}\phi_{2}\phi_{3}\phi_{4}\phi_{5}\phi_{6})$}
The power of deriving amplitudes from a superspace expansion is
that one can rule out certain amplitudes just by inspection.
Amplitudes with more than two scalars carrying the same color
vanishes since there is no way of assigning the correct fermionic
variables. Here we look at six scalar amplitude all carrying
different flavor. This should be the simplest amplitude since the
flavor carried by the internal particle is uniquely determined. We
give the result for cut $t_{123}$ while the other cuts are related
by symmetry.
\begin{eqnarray}
B_{1}(\phi_{1}\phi_{2}\phi_{3}\phi_{4}\phi_{5}\phi_{6})=\{(\langle
12\rangle[56]\langle3|K_{123}|4\rangle+\langle12\rangle[64]\langle3|K_{123}|5\rangle+\langle12\rangle[45]\langle3|K_{123}|6\rangle\\
\nonumber+\langle
31\rangle[56]\langle2|K_{123}|4\rangle+\langle31\rangle[64]\langle2|K_{123}|5\rangle+\langle31\rangle[45]\langle2|K_{123}|6\rangle\\
\nonumber+\langle
23\rangle[56]\langle1|K_{123}|4\rangle+\langle23\rangle[64]\langle1|K_{123}|5\rangle+\langle23\rangle[45]\langle1|K_{123}|6\rangle)^{2}\}B_{0}\\
\nonumber +complex\;conjugate .
\end{eqnarray}

\section{4. A brief discussion on the MHV vertex approach }
As discussed in the introduction, the straight forward way to
compute amplitudes in superspace is the generalization of the MHV
vertex \cite{10} approach. It is also of conceptual interest to
see if this approach actually works for the NMHV loop amplitude.
Here we give a brief discussion of the extension.

The MHV vertex approach was shown to be successful \cite{12} in
constructing the n point MHV loop amplitude. This is partly due to
the similarity between the cut diagrams \cite{1} originally used
to compute the amplitude and the MHV vertex diagram, so that one
can use a dispersion type integral to reconstruct the box
functions from it's discontinuity across the branch cut. For the
NMHV loop amplitude, one requires three propagator for the three
MHV vertex one-particle-irreducible(1PI) diagram and two
propagators for the one-particle-reducible(1PR) diagram
(fig-2)\cite{13}.
\begin{figure}
\includegraphics{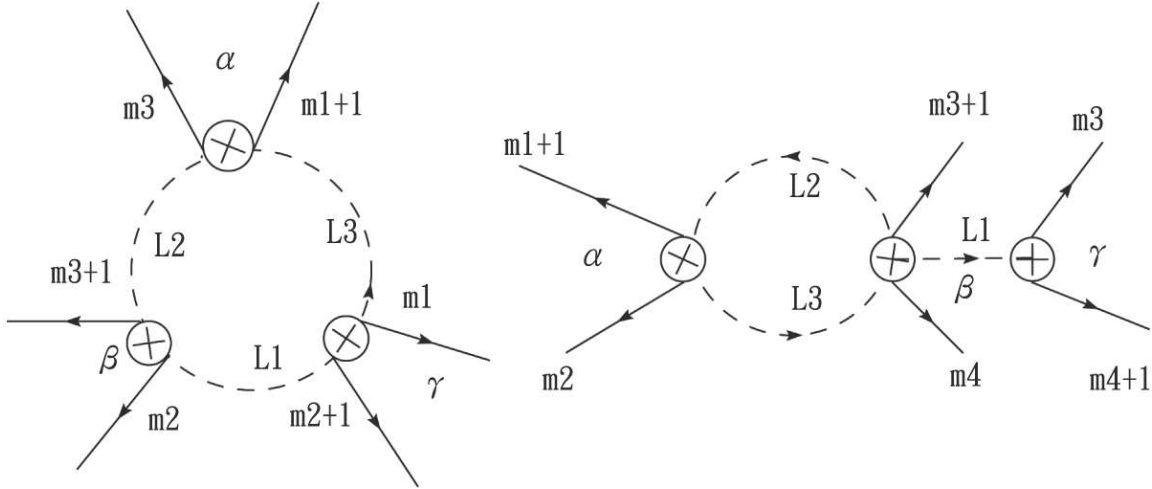}
\caption{\label{fig:2} MHV diagrams for NMHV loop amplitude,
includes the one-particle-irreducible and one-particle-reducible
graph. The $m_{i}$s label external momenta that are adjacent to
the propagator. For the first graph we need to integrate over all
three propagators, while only $L_{2}$ $L_{3}$ are integrated in
the second. One also needs to sum over all possible ways of
assigning external momenta to the vertices}
\end{figure}
We would then encounter the following integration:
\begin{eqnarray}
\frac{1}{\prod_{i=1}^{n}\langle ii+1\rangle}\int\frac{d^{4}L_{1}}{L_{1}^{2}}\frac{d^{4}L_{2}}{L_{2}^{2}}\frac{d^{4}L_{3}}{L_{3}^{2}}\delta(P_{\alpha}+L_{2}-L_{3})\delta(P_{\beta}+L_{3}-L_{1})\delta(P_{\gamma}+L_{1}-L_{2}) \\
\nonumber\int
d^{8}\eta_{l_{1}}d^{8}\eta_{l_{2}}d^{8}\eta_{l_{3}}\frac{\delta^{8}(\Theta_{1})\delta^{8}(\Theta_{2})\delta^{8}(\Theta_{3})\langle
m_{2}m_{2}+1\rangle\langle m_{1}m_{1}+1\rangle\langle
m_{3}m_{3}+1\rangle}{\langle l_{2}l_{1}\rangle \langle
l_{3}l_{2}\rangle\langle l_{1}l_{3}\rangle\langle
l_{1}m_{2}+1\rangle\langle m_{2}l_{1}\rangle\langle
l_{2}m_{3}+1\rangle\langle m_{3}l_{2}\rangle\langle
l_{3}m_{1}+1\rangle\langle m_{1}l_{3}\rangle}\\
\nonumber+\frac{\delta(L_{1}-P_{\gamma})}{\prod_{i=1}^{n}\langle
ii+1\rangle}\int\frac{d^{4}L_{2}}{L_{2}^{2}}\frac{d^{4}L_{3}}{L_{3}^{2}}\delta(P_{\alpha}+L_{3}-L_{2})\delta(P_{\beta}+L_{2}+L_{1}-L_{3})\int
d^{8}\eta_{l_{1}}d^{8}\eta_{l_{2}}d^{8}\eta_{l_{3}}\\
\nonumber\times\frac{\delta^{8}(\Theta_{1})\delta^{8}(\Theta_{2})\delta^{8}(\Theta_{3})\langle
m_{2}m_{2}+1\rangle\langle m_{1}m_{1}+1\rangle\langle
m_{3}m_{3}+1\rangle\langle m_{4}m_{4}+1\rangle}{L_{1}^{2}\langle
l_{3}l_{2}\rangle^{2}\langle m_{1}l_{2}\rangle\langle
l_{2}m_{1+1}\rangle\langle l_{3}m_{2}+1\rangle\langle
m_{2}l_{3}\rangle\langle l_{1}m_{3}+1\rangle\langle
m_{3}l_{1}\rangle\langle l_{1}m_{4}+1\rangle\langle
m_{4}l_{1}\rangle}
\end{eqnarray}
where for the first term
\begin{eqnarray}
\Theta_{1}=\sum_{i=\alpha}\eta_{i}\lambda_{i}+l_{2}\eta_{l_{2}}-l_{3}\eta_{l_{3}}\\
\nonumber
\Theta_{2}=\sum_{i=\beta}\eta_{i}\lambda_{i}+l_{3}\eta_{l_{3}}-l_{1}\eta_{l_{1}}\\
\nonumber\Theta_{3}=\sum_{i=\gamma}\eta_{i}\lambda_{i}+l_{1}\eta_{l_{1}}-l_{2}\eta_{l_{2}}
\end{eqnarray}
for the second term
\begin{eqnarray}
\Theta_{1}=\sum_{i=\alpha}\eta_{i}\lambda_{i}-l_{2}\eta_{l_{2}}+l_{3}\eta_{l_{3}}\\
\nonumber
\Theta_{2}=\sum_{i=\beta}\eta_{i}\lambda_{i}+l_{1}\eta_{l_{1}}+l_{2}\eta_{l_{2}}-l_{3}\eta_{l_{3}}\\
\nonumber\Theta_{3}=\sum_{i=\gamma}\eta_{i}\lambda_{i}-l_{1}\eta_{l_{1}}
\end{eqnarray}
$\alpha$$\beta$$\gamma$ labels the external momenta assigned to
the three MHV vertex and the $l_{i}$s are the off shell
continuation spinor following the CSW prescription \cite{10}. We
can reorganize the delta functions to reproduce the overall
momentum conservation. For the first term in eq.(37) we have
\begin{equation}
\delta(P_{\alpha+\beta+\gamma})\delta(P_{\beta+\gamma}+L_{3}-L_{2})\delta(P_{\gamma}+L_{1}-L_{2})
\end{equation}
For the second term
\begin{equation}
\nonumber\delta(P_{\alpha+\beta+\gamma})\delta(P_{\alpha}+L_{2}-L_{3})
\end{equation}
If we integrate the last delta function away in the first term and
combine with the 1PR graphs, it is equivalent to using two MHV
vertex to construct NMHV tree amplitude on one side of the two
remaining propagator, namely this combines vertex $\gamma$ and
$\beta$ through propagator $L_{1}$. To see this note that the
momentum conserving delta function forces $L_{1}$ propagator to
carry the correct momentum as it would for the CSW method and the
$\frac{1}{P_{L_{1}}^{2}}$ is present in the integral measure in
the first place. This would obviously affect the off shell spinor
in the following way.
\begin{equation}
l_{1}=L_{1}\widetilde{\eta}\rightarrow
(L_{2}-P_{\gamma})\widetilde{\eta}
\end{equation}
This simply fixes the off shell spinor to be computed from the
correct momentum as the CSW method. Thus we have come to a two
propagator integral with two tree level amplitude on both side
constructed from the CSW method. This is exactly the picture one
would have if one apply the standard cut, except the propagators
are off shell instead of on shell. For higher number of MHV
vertices this can be applied straight forward, by integrating the
momentum conserving propagator one at a time one can reduce the
number of propagators until one arrive at the standard cut
picture. As shown in \cite{12} one can then proceed to recast the
two propagator integral into a dispersion integral which computes
the discontinuity across the cut of the integrand, by using the
cut constructibility of N=4 SYM loop amplitudes, one can
reconstruct the box function and it's coefficient. However there
is one subtlety. In the original standard cut one has to analyze
every cut channel, and then disentangle the information since more
than one box integral share the same cuts. If we follow the CSW
prescription we can always reduce the loop diagrams down to two
propagator one loop diagrams with a MHV vertex on one side of the
two propagators. Thus this implies if the CSW approach is valid at
one loop, then the full loop amplitude should be able to be
reconstructed from the cuts of a subgroup of two propagator
diagrams which always have a MHV vertex on one side of the cut.

This construction makes the connection between MHV vertex and
$\overline{MHV}$ loop amplitude more transparent. $\overline{MHV}$
loop are just the parity transformation of the MHV loop, where one
simply take the complex conjugate of the MHV loop:
\begin{equation}
A(\overline{MHV})_{loop}=A(\overline{MHV})_{tree}\sum_{i=1}^{n}\sum_{r=1}^{[n/2]-1}(1-\frac{1}{2}\delta_{\frac{n}{2}},r)F^{2me}_{n:r;i}
\end{equation}
It's derivation from MHV vertex is as follows. In \cite{15} it was
shown that using MHV vertices one can reconstruct the
$\overline{MHV}$ tree amplitude in it's complex conjugate spinor
form. Since by integrating out one loop propagator corresponds to
using MHV vertex to construct NMHV tree amplitude, one can proceed
in a specific manner to reduce the number of loop propagators down
to two with two $\overline{MHV}$ tree on both side. Since from
\cite{15} the two $\overline{MHV}$ tree amplitudes on both side is
expressed in complex conjugate form, following exactly the same
lines in \cite{12} one can reproduce eq.(43).
\section{5. Conclusion}
  In this paper we constructed the 6 point NMHV loop amplitude for
  N=4 SuperYang-Mills in a compact form using its cut constructible nature. The expansion with
  respect to the fermionic parameter gives amplitudes with
  different particle content and helicity ordering. To extend
  further to higher point NMHV loops one may have to resolve to
  the MHV vertex approach since the tree level amplitudes on both
  side of the cut in general will not be in simple MHV and
  $\overline{MHV}$ combination. We also give a general discussion on how to proceed with the MHV
vertex construction for higher than MHV loop(more than two
negative helicities). The fact that it reproduces the two
propagator picture for any one loop diagram combined with earlier
results that have reproduced the MHV loop\cite{12} and the
relationship between the leading order and sub leading order
amplitudes\cite{13}, gives a strong support for the CSW approach
beyond tree level.
\section{6. Acknowledgements}
It is a great pleasure to thank Warren Siegel for suggesting the
question, Gabriele Travaglini, L. J. Dixon and Freddy Cachazo for
their useful comments.


\begin{thebibliography}{99}
\bibitem{1}
  Z.~Bern, L.~J.~Dixon, D.~C.~Dunbar and D.~A.~Kosower,
  Nucl.\ Phys.\ B {\bf 425}, 217 (1994)
  [arXiv:hep-ph/9403226].
  \bibitem{2}
  Z.~Bern, L.~J.~Dixon, D.~C.~Dunbar and D.~A.~Kosower,
  Nucl.\ Phys.\ B {\bf 435}, 59 (1995)
  [arXiv:hep-ph/9409265].
\bibitem{3}
  R.~Britto, F.~Cachazo and B.~Feng,
  arXiv:hep-th/0412103.
\bibitem{4}
Z.~Bern, L.~J.~Dixon and D.~A.~Kosower,
  Nucl.\ Phys.\ B {\bf 513}, 3 (1998)
  [arXiv:hep-ph/9708239].
 Z.~Bern, L.~J.~Dixon and D.~A.~Kosower,
  JHEP {\bf 0408}, 012 (2004)
  [arXiv:hep-ph/0404293].

 Z.~Bern, V.~Del Duca, L.~J.~Dixon and D.~A.~Kosower,
  Phys.\ Rev.\ D {\bf 71}, 045006 (2005)
  [arXiv:hep-th/0410224].


\bibitem{5}
  S.~J.~Bidder, D.~C.~Dunbar and W.~B.~Perkins,
  arXiv:hep-th/0505249.
\bibitem{6}
  M.~T.~Grisaru, H.~N.~Pendleton and P.~van Nieuwenhuizen,
  Phys.\ Rev.\ D {\bf 15}, 996 (1977).
\bibitem{7}
  V.~P.~Nair,
  Phys.\ Lett.\ B {\bf 214}, 215 (1988).
\bibitem{8}
  E.~Witten,
  Commun.\ Math.\ Phys.\  {\bf 252}, 189 (2004)
  [arXiv:hep-th/0312171].
\bibitem{9}
  R.~Britto, F.~Cachazo and B.~Feng,
  Nucl.\ Phys.\ B {\bf 715}, 499 (2005)
  [arXiv:hep-th/0412308].

  R.~Britto, B.~Feng, R.~Roiban, M.~Spradlin and A.~Volovich,
  Phys.\ Rev.\ D {\bf 71}, 105017 (2005)
  [arXiv:hep-th/0503198].

  R.~Britto, F.~Cachazo, B.~Feng and E.~Witten,
  Phys.\ Rev.\ Lett.\  {\bf 94}, 181602 (2005)
  [arXiv:hep-th/0501052].

  \bibitem{10}
  F.~Cachazo, P.~Svrcek and E.~Witten,
  JHEP {\bf 0409}, 006 (2004)
  [arXiv:hep-th/0403047].

  \bibitem{11}
  G.~Georgiou, E.~W.~N.~Glover and V.~V.~Khoze,
  JHEP {\bf 0407}, 048 (2004)
  [arXiv:hep-th/0407027].

\bibitem{12}
  A.~Brandhuber, B.~Spence and G.~Travaglini,
  Nucl.\ Phys.\ B {\bf 706}, 150 (2005)
  [arXiv:hep-th/0407214].

\bibitem{13}
  M.~x.~Luo and C.~k.~Wen,
  Phys.\ Lett.\ B {\bf 609}, 86 (2005)
  [arXiv:hep-th/0410118].

  M.~x.~Luo and C.~k.~Wen,
  JHEP {\bf 0411}, 004 (2004)
  [arXiv:hep-th/0410045].

\bibitem{14}
  G.~Georgiou, E.~W.~N.~Glover and V.~V.~Khoze,
  JHEP {\bf 0407}, 048 (2004)
  [arXiv:hep-th/0407027].


\bibitem{15}
  J.~B.~Wu and C.~J.~Zhu,
  JHEP {\bf 0407}, 032 (2004)
  [arXiv:hep-th/0406085].

 J.~B.~Wu and C.~J.~Zhu,
  JHEP {\bf 0409}, 063 (2004)
  [arXiv:hep-th/0406146].

\end{thebibliography}
\end{document}